\documentclass[aps,prc,twocolumn,nofootinbib,superscriptaddress,floatfix,10pt]{revtex4-2}

\usepackage{amsmath}
\usepackage{amssymb}
\usepackage{bm}
\usepackage{graphicx}
\usepackage{xcolor}
\usepackage[utf8]{inputenc}
\usepackage[T1]{fontenc}
\usepackage{microtype}
\usepackage{booktabs}
\usepackage{dcolumn}
\usepackage{hyperref}
\usepackage{lineno} 

\hypersetup{
    colorlinks=true,
    linkcolor=blue,
    citecolor=blue,
    urlcolor=blue
}

\newcolumntype{d}[1]{D{.}{.}{#1}}

\begin{document}

\title{Multi-Energy Analysis of Baryon Stopping: Extracting Non-Equilibrium Dynamics via Stochastic Rapidity Diffusion}

\author{Dhananjaya Thakur}
\email{dhananjaya.phy@itbhu.ac.in}
\affiliation{Department of Physics, Indian Institute of Technology (BHU), Varanasi 221005, India}

\date{October 21, 2026}

\pacs{25.75.Dw, 25.75.Nq, 12.38.Mh}

\begin{abstract}
We present a comprehensive extraction of baryon stopping transport properties derived from net-proton transverse momentum ($p_T$) spectra across a broad energy range, from $E_{\mathrm{lab}} = 2.0\,A\text{ GeV}$ ($\sqrt{s_{\mathrm{NN}}} = 2.70\text{ GeV}$) to $\sqrt{s_{\mathrm{NN}}} = 62.4\text{ GeV}$ in central heavy-ion collisions ($\le 10\%$ centrality). By performing a systematic, multi-rapidity blast-wave analysis of experimental invariant yields, we extract the continuous rapidity density profile $\rho_{\mathrm{net}}(y)$ and the transverse momentum broadening profile $\langle p_T^2 \rangle_{\mathrm{net}}(y)$. Utilizing a stochastic diffusion framework governed by relativistic Fokker-Planck dynamics, we analytically derive the longitudinal friction coefficient $\alpha_{\mathrm{bw}}$, the rapidity diffusion coefficient $D_y$, the longitudinal relaxation time $\tau_{\mathrm{relax}}$, and the associated longitudinal Reynolds number $\mathrm{Re}_L$. Our extracted central stopping fractions demonstrate qualitative consistency with the previous published two source thermal calculations. Crucially, the excitation function of $\mathrm{Re}_L$ reveals a drop by a factor of $\sim 163$, from $\mathrm{Re}_L \approx 4.88$ at $\sqrt{s_{\mathrm{NN}}} = 2.70\text{ GeV}$ to $\mathrm{Re}_L \approx 0.03$ at $62.4\text{ GeV}$. This provides a clear macroscopic baseline for the crossover
from inertial, advection-dominated baryon stopping to thermal, diffusion-dominated nuclear transparency, establishing a solid data-driven baseline for the topological baryon junction dynamics and high-$\mu_B$ matter anticipated at upcoming facilities, including FAIR, NICA, and the CEE.
\end{abstract}

\maketitle

\section{INTRODUCTION}
\label{sec:intro}

Understanding the early-stage dynamics of ultrarelativistic heavy-ion collisions is crucial for mapping the Quantum Chromodynamics (QCD) phase diagram~\cite{stephanov1998,ratti2021}. A key observable sensitive to these dynamics is baryon stopping---the deceleration of incident nucleons in the longitudinal direction. Because baryon number is a strictly conserved quantum charge, tracking the transport of net-baryons (typically proxied by net-protons) from beam rapidities toward mid-rapidity provides an explicit record of the energy deposition during the earliest moments of the collision~\cite{Wong:2000cp, wolschin2007, Thakur:2016znw, Rode:2026hxh}. The net-proton rapidity distribution and its transverse momentum fluctuations provide direct insights into the friction and diffusion processes governing the hot, dense medium created in the collision zone~\cite{bass1998, ivanov2017}.

Recently, the STAR Collaboration at the Relativistic Heavy Ion Collider (RHIC) has provided definitive microscopic evidence challenging the conventional picture of baryon number transport~\cite{star2026}. Specifically, by measuring the net-baryon to net-charge ratio ($B/\Delta Q$) at mid-rapidity in isobar collisions ($^{96}_{44}\mathrm{Ru}+^{96}_{44}\mathrm{Ru}$ vs. $^{96}_{40}\mathrm{Zr}+^{96}_{40}\mathrm{Zr}$), STAR observed a value far exceeding the naive valence quark expectation, indicating that baryon number is transported independently of electric charge. Furthermore, by simulating the rapidity dependence of net-proton yields in photonuclear ($\gamma + \mathrm{Au}$) interactions, they found a significantly less asymmetric distribution than predicted by valence quark transport models. These groundbreaking measurements provide the definitive microscopic proof that baryon number is not carried primarily by fractionally charged valence quarks, but rather by the non-perturbative topological baryon junction---a Y-shaped configuration of neutral gluon fields~\cite{rossi1977,artru1981,kharzeev1996,vance1998,kopeliovich1999}.

This paradigm shift is central to our theoretical framework. Conventionally, the deceleration of valence quarks through multiple scatterings with soft gluons has been used to explain stopping ~\cite{Mehtar-Tani:2008hpn}. However, these models struggle to account for the high mid-rapidity baryon densities observed at low energies and the significant transparency at high energies. In stark contrast, the baryon junction paradigm~\cite{kharzeev1996,vance1998,kopeliovich1999} posits that the Y-shaped topological junction carries the baryon number. Because this junction carries a minuscule fraction of the nucleon's longitudinal momentum, it is highly susceptible to catastrophic momentum loss in dense gluon fields. This leads to enhanced baryon densities at mid-rapidity and a characteristic exponential drop in the stopping cross-section with rapidity loss~\cite{kopeliovich1999}. The transition from a fully stopped, baryon-rich environment at AGS energies to a transparent, baryon-depleted mid-rapidity at SPS/RHIC energies may correspond to chiral symmetry restoration, allowing the junction to decouple independently from the color fields~\cite{stephanov1998,fodor2004,aoki2006,bazavov2012}.

Historically, the two analytical domains of bulk thermodynamics and longitudinal transport have been treated in relative isolation. Transverse momentum ($p_T$) spectra are routinely analyzed using static, hydrodynamically inspired models~\cite{schnedermann1993,heinz2001,retiere2004}, which effectively captures the kinetic freeze-out temperature and radial expansion velocity. Concurrently, the longitudinal rapidity distributions of net-baryons are frequently analyzed through phenomenological models, including the relativistic diffusion model~\cite{Denicol:2018wdp}, which formulates baryon transport as a stochastic Markov process in phase space governed by Fokker-Planck equations~\cite{risken1989,gardiner2004}.

The primary limitation in the existing literature is the lack of an automated, integrated framework capable of simultaneously fitting multi-energy invariant yields, deriving local variance scalings, and extracting unified macroscopic transport variables (such as the rapidity diffusion coefficient $D_y$, the longitudinal Reynolds number $\mathrm{Re}_L$, and entropy production) with coupled systematic error propagation. The computational framework evaluated in this work resolves this gap. By utilizing an integration over invariant variables, bounded by strict physical limits, the framework provides an end-to-end analysis pipeline. We develop a data analysis framework that processes net-proton spectra across six distinct beam energies ($E_{\mathrm{lab}} = 2.0, 4.0, 6.0, 8.0, 158.0\,A\text{ GeV}$, and $\sqrt{s_{\mathrm{NN}}} = 62.4\text{ GeV}$) in central collisions. By fitting invariant yields with a blast-wave model and extracting the rapidity-dependent density and variance profiles, we utilize a stochastic diffusion framework to directly extract the friction and diffusion coefficients governing the longitudinal stopping process.

The subsequent sections detail the theoretical formulation of the integrated model, evaluate computational stability, present the extracted transport dynamics across six different collision energies, and provide a comparative cross-validation against our previously published stopping fraction numbers~\cite{thakur2016}.

\section{THEORETICAL FRAMEWORK AND DATA ANALYSIS}
\label{sec:theory}

Natural units $\hbar = c = k_B = 1$ are employed throughout this work unless explicitly stated. The framework is constructed upon two primary theoretical pillars: the invariant kinematics of the expanding fireball at kinetic freeze-out, and the non-equilibrium stochastic diffusion describing the longitudinal deceleration of baryon numbers. By linking these domains, the model extracts a complete thermodynamic and transport profile of the collision.

\subsection{Invariant Blast-Wave Kinematics and Transverse Expansion}

For the analysis in this work, the input data consist of experimental $p_T$ spectra for protons ($p$) and anti-protons ($\bar{p}$) across multiple rapidity bins in central collisions ($\le 10\%$ centrality). Net-protons serve as a reliable experimental proxy for net-baryon transport; while hyperon decays ($\Lambda \to p\pi^-$) and isospin differences between protons and neutrons contribute to absolute normalizations, isospin fluctuations largely average out at mid-rapidity in symmetric nucleus-nucleus collisions, ensuring that net-proton phase-space gradients accurately reflect net-baryon deceleration~\cite{Videbaek:2005qn}. The net-proton differential yield is calculated as:
\begin{equation}
\frac{\mathrm{d}^2N_{\mathrm{net}}}{\mathrm{d}y\,\mathrm{d}p_T} = \frac{\mathrm{d}^2N_p}{\mathrm{d}y\,\mathrm{d}p_T} - \frac{\mathrm{d}^2N_{\bar{p}}}{\mathrm{d}y\,\mathrm{d}p_T}.
\label{eq:net_yield}
\end{equation}

The colliding frame kinematics are defined by the proton rest mass $m_p = 0.938272\text{ GeV}$. For each beam energy $E_{\mathrm{lab}}$ (in GeV per nucleon), the invariant center-of-mass energy and beam rapidity are given by:
\begin{align}
\sqrt{s_{\mathrm{NN}}} &= \sqrt{2m_p\left(E_{\mathrm{lab}} + m_p\right) + 2m_p^2}, \label{eq:sqrt_snn} \\
y_{\mathrm{beam}} &= \operatorname{arcosh}\left(\frac{\sqrt{s_{\mathrm{NN}}}}{2m_p}\right). \label{eq:y_beam}
\end{align}
For fixed-target measurements reported in the laboratory system, rapidities are transformed into the center-of-mass frame via $y_{\mathrm{CM}} = y_{\mathrm{lab}} - y_{\mathrm{cms,lab}}$, where $y_{\mathrm{cms,lab}} = \frac{1}{2}\operatorname{arcosh}((E_{\mathrm{lab}}+m_p)/m_p)$.

The invariant blast-wave model, originally formulated by Schnedermann, Sollfrank, and Heinz (SSH)~\cite{schnedermann1993}, assumes a thermalized, longitudinally boost-invariant source undergoing collective radial expansion. The invariant spectral density is given by the Cooper-Frye formula integrated over the freeze-out hypersurface $\Sigma$~\cite{Cooper:1974mv}:
\begin{equation}
E\frac{\mathrm{d}^3N}{\mathrm{d}^3p} = \frac{g}{(2\pi)^3}\int_{\Sigma} f(x,p)\, p^\mu \mathrm{d}\Sigma_\mu,
\label{eq:cooper_frye}
\end{equation}
where $f(x,p) = \exp(-p^\mu u_\mu / T_{\mathrm{kin}})$ is the local equilibrium Boltzmann distribution at kinetic freeze-out temperature $T_{\mathrm{kin}}$, and $g = 2$ is the nucleon spin degeneracy factor.

Under the assumption of a longitudinally boost-invariant Bjorken expansion, the freeze-out hypersurface is characterized by constant proper time $\tau_f$ and space-time rapidity $\eta$, with volume element $\mathrm{d}\Sigma_\mu = (\tau_f \cosh\eta, 0, 0, -\tau_f \sinh\eta)\,r\,\mathrm{d}r\,\mathrm{d}\eta\,\mathrm{d}\phi_b$. In this framework, the particle four-momentum is expressed in terms of transverse mass $m_T = \sqrt{p_T^2 + m_p^2}$, rapidity $y$, and azimuthal angle $\phi$:
\begin{equation}
p^\mu = (m_T \cosh y,\, p_T \cos\phi,\, p_T \sin\phi,\, m_T \sinh y).
\label{eq:four_momentum}
\end{equation}
The fluid velocity profile is parameterized by a linear radial expansion velocity $\beta(r) = \beta_s(r/R)$, corresponding to flow rapidity $\rho(r) = \operatorname{artanh}\beta(r)$. The fluid four-velocity is given by:
\begin{equation}
u^\mu = (\cosh\rho \cosh\eta,\, \sinh\rho \cos\phi_b,\, \sinh\rho \sin\phi_b,\, \cosh\rho \sinh\eta).
\end{equation}

The scalar product $p^\mu u_\mu$ becomes:
\begin{equation}
p^\mu u_\mu = m_T \cosh(y-\eta)\cosh\rho - p_T \sinh\rho \cos(\phi-\phi_b).
\label{eq:product}
\end{equation}

Substituting Eq.~\eqref{eq:product} into Eq.~\eqref{eq:cooper_frye} and integrating over the azimuthal angle $\phi$, the modified Bessel functions $I_0$ and $K_1$ emerge analytically~\cite{schnedermann1993}. The final invariant spectral density for particles emitted from the thermalized fireball is:
\begin{equation}
\begin{split}
\frac{\mathrm{d}^2N}{\mathrm{d}y\,\mathrm{d}p_T} = \mathcal{N}p_T \int_{-\eta_{\mathrm{max}}}^{\eta_{\mathrm{max}}} & m_T \cosh(y-\eta)\cosh\rho \\
& \times I_0\left(\frac{p_T\sinh\rho}{T_{\mathrm{kin}}}\right) \\
& \times K_1\left(\frac{m_T\cosh(y-\eta)\cosh\rho}{T_{\mathrm{kin}}}\right)\mathrm{d}\eta,
\end{split}
\label{eq:invariant_bw}
\end{equation}
where $\mathcal{N} = \frac{g\tau_f}{(2\pi)^2}\int r\,\mathrm{d}r$ is an overall normalization parameter, and $\eta_{\mathrm{max}} = y_{\mathrm{max}} + 0.6$ ensures complete coverage of the longitudinal source profile without truncation artifacts.

To evaluate systematic uncertainties, fits are conducted over three independent $p_T$ windows: $0.35 \le p_T \le 3.0\text{ GeV}/c$ (nominal), $0.30 \le p_T \le 2.5\text{ GeV}/c$ (Sys1), and $0.40 \le p_T \le 3.2\text{ GeV}/c$ (Sys2)~\cite{heinz2001,retiere2004}. From these fits, we extract the normalized rapidity density:
\begin{equation}
\rho_{\mathrm{net}}(y) = \int_0^\infty \frac{\mathrm{d}^2N_{\mathrm{net}}}{\mathrm{d}y\,\mathrm{d}p_T}\,\mathrm{d}p_T
\label{eq:rho}
\end{equation}
and the transverse momentum broadening:
\begin{equation}
\langle p_T^2\rangle_{\mathrm{net}}(y) = \frac{1}{\rho_{\mathrm{net}}(y)}\int_0^\infty p_T^2\,\frac{\mathrm{d}^2N_{\mathrm{net}}}{\mathrm{d}y\,\mathrm{d}p_T}\,\mathrm{d}p_T.
\label{eq:pt2}
\end{equation}

In heavy-ion collisions, a fraction of observed protons and antiprotons originates from the decay of heavier baryonic resonances (such as $\Delta(1232) \to N + \pi$, $N(1535) \to N + \pi$) and hyperon feed-down~\cite{alice2013, bearden2004}. Our framework treats the blast-wave model as an effective parameterization of the inclusive spectral shape. When fitted to inclusive yields, the SSH functional form integrates over the combined spectral distribution, yielding an effective kinetic freeze-out temperature ($T_{\mathrm{kin}}$) and radial flow velocity ($\beta$) that describe the freeze-out surface of the observed particle population~\cite{schnedermann1993,retiere2004}.

These discrete profiles are subsequently parameterized across rapidity. The rapidity density is modeled by a symmetric double-Gaussian function~\cite{ivanov2017}:
\begin{equation}
\rho_{\mathrm{net}}(y) = A\left[\exp\left(-\frac{(y - y_{\mathrm{shift}})^2}{2w^2}\right) + \exp\left(-\frac{(y + y_{\mathrm{shift}})^2}{2w^2}\right)\right],
\label{eq:rho_profile}
\end{equation}
while the variance follows a $\operatorname{sech}^2$ scaling~\cite{retiere2004}:
\begin{equation}
\langle p_T^2\rangle_{\mathrm{net}}(y) = C_0 + \frac{C_1}{\cosh^2(y/C_2)}.
\label{eq:pt2_profile}
\end{equation}

Here, $A$ is the normalization amplitude, $y_{\mathrm{shift}}$ denotes the mean rapidity of decelerated fragmentation peaks, and $w$ represents rapidity broadening. In Eq.~\eqref{eq:pt2_profile}, $C_0$ establishes the non-flow transverse variance baseline, which physically absorbs the kinematic broadening caused by the decay of heavier baryonic resonances and strange hyperons. This successfully isolates the true longitudinal-transverse diffusion coupling. $C_1$ quantifies the mid-rapidity kinetic enhancement, and $C_2$ dictates the characteristic rapidity scale of transverse flow~\cite{heinz2001}.

\subsection{Numerical Integration and Parameter Bounding}

Evaluation of the invariant blast-wave integral can encounter numerical instabilities due to the asymptotic behavior of $K_1(x)$ at small arguments and rapid hyperbolic growth at large rapidities~\cite{schnedermann1993,retiere2004}. The computational pipeline stabilizes execution by enforcing parameter boundaries. Specifically, bounding the maximum rapidity to the limit plus an offset of $\eta_{\mathrm{max}} = y_{\mathrm{max}} + 0.6$ and setting a minimum kinetic temperature of $T_{\mathrm{kin}} \ge 0.030\text{ GeV}$ prevents asymptotic divergence of the modified Bessel function while safely capturing the thermal Gaussian tail.

To maintain robust convergence across energies spanning three orders of magnitude, dynamic logarithmic initializations are applied:
\begin{align}
T_{\mathrm{seed}} &= 0.100 + 0.003 \ln(\sqrt{s_{\mathrm{NN}}}/2.5) \quad [\text{GeV}], \label{eq:T_seed} \\
\beta_{\mathrm{seed}} &= 0.300 + 0.040 \ln(\sqrt{s_{\mathrm{NN}}}/2.5), \label{eq:beta_seed}
\end{align}
with upper ceilings bounded at $T_{\mathrm{seed}} \le 0.138\text{ GeV}$ and $\beta_{\mathrm{seed}} \le 0.580$. This dynamic seeding provides numerical stability across the entire beam energy scan.

\subsection{Systematic Error Quantification}

Uncertainties in extracted freeze-out parameters are evaluated by combining statistical errors from the fit covariance matrix with systematic variations across the $p_T$ windows:
\begin{equation}
\sigma_{X}^{\mathrm{sys}} = \frac{1}{2}\left(|X_{\mathrm{nom}} - X_{\mathrm{sys1}}| + |X_{\mathrm{nom}} - X_{\mathrm{sys2}}|\right),
\label{eq:sys}
\end{equation}
where $X \in \{T_{\mathrm{kin}}, \beta\}$. The total uncertainty is added in quadrature:
\begin{equation}
\sigma_{X}^{\mathrm{final}} = \sqrt{(\sigma_{X}^{\mathrm{stat}})^2 + (\sigma_{X}^{\mathrm{sys}})^2}.
\label{eq:total_err}
\end{equation}

Uncertainties are propagated downstream into transport observables: an $8\%$ relative systematic uncertainty is assigned to the friction coefficient $\alpha_{\mathrm{bw}}$ for numerical gradient evaluations; the uncertainty on the diffusion coefficient $D_y$ is propagated in quadrature from $\alpha_{\mathrm{bw}}$ and a $20\%$ uncertainty on the input momentum diffusion constant ($D_{p_T} = 0.08\text{ GeV}^2/\text{fm}$, $\delta D_{p_T} = 0.016\text{ GeV}^2/\text{fm}$); the relaxation time $\tau_{\mathrm{relax}}$ inherits the fractional uncertainty of $\alpha_{\mathrm{bw}}$; and the Reynolds number $\mathrm{Re}_L$, entropy production rate $\mathrm{d}S_B/(\mathrm{d}y\,\mathrm{d}\tau)$, and stopping fraction $f_{\mathrm{stop}}$ are evaluated with baseline relative uncertainties of $9\%$, $12\%$, and $5\%$, respectively.

\subsection{Stochastic Transport and Fokker-Planck Dynamics}

Longitudinal baryon transport is governed by a stochastic diffusion process in rapidity space described via the forward Fokker-Planck equation (FPE) for the distribution function $P(y,\tau)$~\cite{wolschin2007,risken1989}:
\begin{equation}
\frac{\partial P(y,\tau)}{\partial \tau} = -\frac{\partial}{\partial y}\left[ K_1(y)P(y,\tau) \right] + \frac{\partial^2}{\partial y^2}\left[ K_2(y)P(y,\tau) \right],
\label{eq:fpe}
\end{equation}
where $K_1(y)$ is the drift vector driving rapidity loss, $K_2(y) = D_y$ is the rapidity diffusion coefficient, and $\tau$ is proper time.

Linearizing the restorative drift force toward mid-rapidity, $K_1(y) = -\alpha y$, casts the evolution into an Ornstein-Uhlenbeck process~\cite{gardiner2004}. Equation~\eqref{eq:alpha} follows directly from combining this Fokker-Planck continuity equation with the diffusion approximation in $p_T^2$ space. Taking the second transverse momentum moment of the stationary distribution and coupling the local transverse energy variance density $W(y) = \rho_{\mathrm{net}}(y)\langle p_T^2 \rangle_{\mathrm{net}}(y)$ to the accumulating net-baryon flux $\Phi(y) = \int_{0}^{y} \rho_{\mathrm{net}}(y')\,\mathrm{d}y'$ yields the force-balance condition: $\alpha_{\mathrm{bw}}\cdot \Phi(y) = -\partial W(y)/\partial y$. Expanding via the product rule and evaluating at $y_{\mathrm{eval}}$ yields:
\begin{equation}
\begin{split}
\alpha_{\mathrm{bw}} = \left| \frac{\rho_{\mathrm{net}}(y_{\mathrm{eval}})}{\Phi(y_{\mathrm{eval}})} \Bigg[ \right. & \left.\frac{\mathrm{d}\langle p_T^2\rangle_{\mathrm{net}}}{\mathrm{d}y}\right|_{y_{\mathrm{eval}}} \\
& \left. + \left.\frac{1}{\rho_{\mathrm{net}}(y_{\mathrm{eval}})} \frac{\mathrm{d}\rho_{\mathrm{net}}}{\mathrm{d}y}\right|_{y_{\mathrm{eval}}} \langle p_T^2\rangle_{\mathrm{net}}(y_{\mathrm{eval}}) \Bigg] \right|,
\end{split}
\label{eq:alpha}
\end{equation}
where the absolute value enforces $\alpha_{\mathrm{bw}} > 0$ for the decelerating net-baryon current.

Physically, the choice of evaluation coordinate $y_{\mathrm{eval}}$ tracks the topological transition of the nuclear collision fireball. At low collision energies ($\sqrt{s_{\mathrm{NN}}} \le 10.0\text{ GeV}$), heavy-ion systems undergo full nuclear stopping, creating a single merged fireball where the bulk gradient is optimally characterized at $y_{\mathrm{eval}} = 0.50$. In contrast, at high collision energies ($\sqrt{s_{\mathrm{NN}}} > 15.0\text{ GeV}$), the system enters the transparent regime where the projectile and target fragmentation peaks recede toward forward and backward rapidities ($\Delta y = y_{\mathrm{beam}} - y_{\mathrm{shift}} \gtrsim 1.0$). In this transparent regime, evaluating at $y_{\mathrm{eval}} = 0.20$ cleanly isolates central stochastic diffusion and topological junction transport from the receding fragmentation remnants.

Microscopically, single-particle momentum dissipation follows a relativistic Langevin equation~\cite{gardiner2004,wolschin2007,DunkelHanggi2009}:
\begin{equation}
\frac{\mathrm{d}p_z}{\mathrm{d}\tau} = -\gamma p_z + \xi(\tau),
\label{eq:langevin}
\end{equation}
where $\gamma$ is the microscopic drag coefficient, $\tau$ is proper time, and the stochastic noise satisfies $\langle \xi(\tau)\xi(\tau') \rangle = 2 D_{p_T}\delta(\tau-\tau')$. Near mid-rapidity ($p_z \to 0$, $E \to m_T$), the phase-space differential maps as $\mathrm{d}y \approx \mathrm{d}p_z/m_T$. Integrating the stochastic noise variance across the isotropic $4\pi$ solid angle introduces a phase-space factor of $\xi = 4$ for forward-backward symmetry. Applying the generalized Einstein fluctuation-dissipation theorem yields:
\begin{equation}
D_y = \frac{4 D_{p_T}}{\alpha_{\mathrm{bw}} \cdot m_p^2},
\label{eq:Dy}
\end{equation}
where $D_{p_T} = 0.08\text{ GeV}^2/\text{fm}$ ($\pm 20\%$) is a standard phenomenological estimate for the transverse momentum diffusion coefficient, consistent with perturbative QCD calculations of heavy-quark transport in the QGP~\cite{Moore:2004tg}, and the nucleon mass squared $m_p^2 \approx 0.88\text{ GeV}^2 \approx 1.0\text{ GeV}^2$ supplies the natural dimensional conversion from transverse momentum squared dispersion to dimensionless rapidity variance.

The longitudinal relaxation time $\tau_{\mathrm{relax}}$, representing the characteristic deceleration time of the incoming protons, is: 
\begin{equation}
\tau_{\mathrm{relax}} = \frac{1}{\alpha_{\mathrm{bw}}} \quad [\text{fm}/c].
\label{eq:tau}
\end{equation}

The dimensionless longitudinal Reynolds number $\mathrm{Re}_L$, characterizing the ratio of inertial advection to diffusive transport, is defined as:
\begin{equation}
\mathrm{Re}_L = \frac{V_{\mathrm{adv}}(y_{\mathrm{eval}})}{D_y} = \frac{\alpha_{\mathrm{bw}} y_{\mathrm{eval}}}{D_y}.
\label{eq:Reynolds}
\end{equation}

From linear irreversible thermodynamics, the local entropy production rate \(\sigma\) is given by the bilinear product of dissipative thermodynamic fluxes \(J_i\) and their conjugate forces \(X_i\), i.e., \(\sigma = \sum_i J_i X_i\).

In rapidity space, the net-baryon diffusive flux is given by Fick's law~\cite{fick1855}:
\begin{equation}
J_y = -D_y \frac{\partial \rho_{\mathrm{net}}}{\partial y} = -D_y \rho_{\mathrm{net}} \frac{\partial \ln \rho_{\mathrm{net}}}{\partial y}.
\end{equation}

The conjugate thermodynamic force driving this diffusion is the gradient of the chemical potential over temperature, \(X_y = -\frac{\partial}{\partial y}\left(\frac{\mu}{T}\right)\). For a dilute system governed by the ideal gas equation of state, the Gibbs-Duhem relation yields \(\mathrm{d}(\mu/T) = \mathrm{d}\ln \rho_{\mathrm{net}}\). Thus, the thermodynamic force is simply the logarithmic density gradient:
\begin{equation}
X_y = -\frac{\partial \ln \rho_{\mathrm{net}}}{\partial y}.
\end{equation}

Multiplying the flux by the force (\(J_y \times X_y\)) cancels the negative signs and yields the non-equilibrium net-baryon entropy production rate per unit rapidity per unit proper time \cite{koide2007}
\begin{equation}
\left. \frac{\mathrm{d}S_B}{\mathrm{d}y\mathrm{d}\tau} \right|_{y_{\mathrm{eval}}} = D_y \cdot \rho_{\mathrm{net}}(y_{\mathrm{eval}}) \cdot \left( \left. \frac{\mathrm{d}\ln \rho_{\mathrm{net}}}{\mathrm{d}y} \right|_{y_{\mathrm{eval}}} \right)^2,
\end{equation}

having natural dimensions of $\text{fm}^{-1}$.

\section{RESULTS AND DISCUSSION}
\label{sec:result}

\subsection{Transverse Momentum Spectra and Blast-Wave Fits}

The invariant transverse momentum ($p_T$) spectra of protons and anti-protons across rapidities and energies are fitted with the SSH blast-wave model. Datasets encompass central collisions from AGS ($E_{\mathrm{lab}} = 2.0, 4.0, 6.0, 8.0\,A\text{ GeV}$)~\cite{E895:2001zms}, SPS ($E_{\mathrm{lab}} = 158\,A\text{ GeV}$)~\cite{NA49:1998gaz}, and RHIC ($\sqrt{s_{\mathrm{NN}}} = 62.4\text{ GeV}$)~\cite{BRAHMS:2009wlg}. Yields are converted to a standardized $\mathrm{d}^2N/(\mathrm{d}y\,\mathrm{d}p_T)$ format.  

\begin{figure}[t]
\centering
\includegraphics[width=\columnwidth]{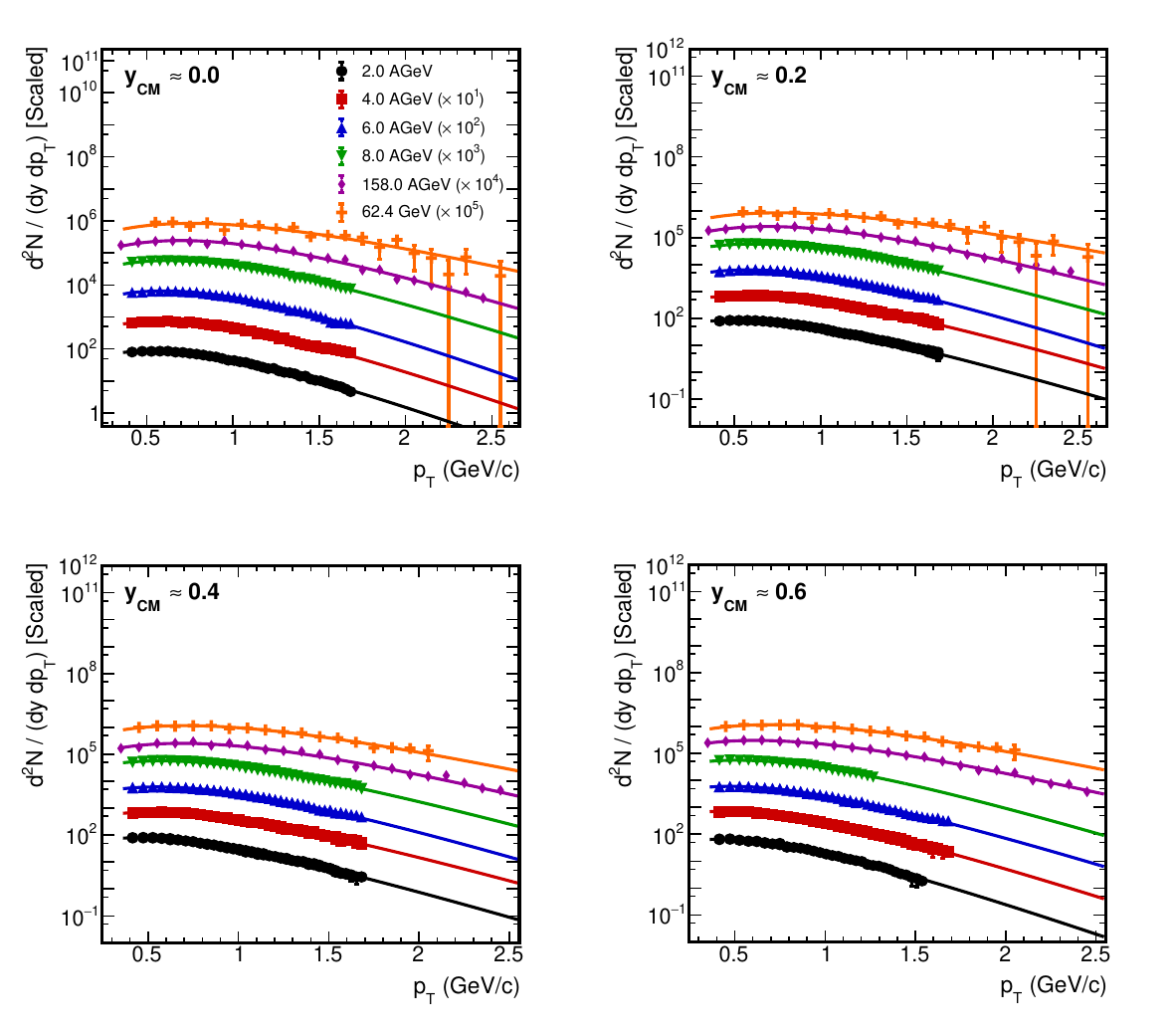}
\caption{\textbf{Transverse momentum spectra of net-protons across rapidity slices.} Scaled invariant yields $\mathrm{d}^2N/(\mathrm{d}y\,\mathrm{d}p_T)$ are shown for rapidities $y_{\mathrm{CM}} \approx 0.0, 0.2, 0.4$, and $0.6$ across all collision energies. Yields are scaled by powers of ten for visual separation. Closed markers represent experimental data; solid curves denote the SSH blast-wave fits.}
\label{fig:pt_spectra}
\end{figure}

The resulting fits across representative rapidity slices are shown in Fig.~\ref{fig:pt_spectra}. Across all energies and rapidities, the blast-wave parameterizations reproduce experimental yields within acceptable $\chi^2/\mathrm{ndf}$.

\subsection{Rapidity Density Distributions}

\begin{figure*}[t]
\centering
\includegraphics[width=\textwidth]{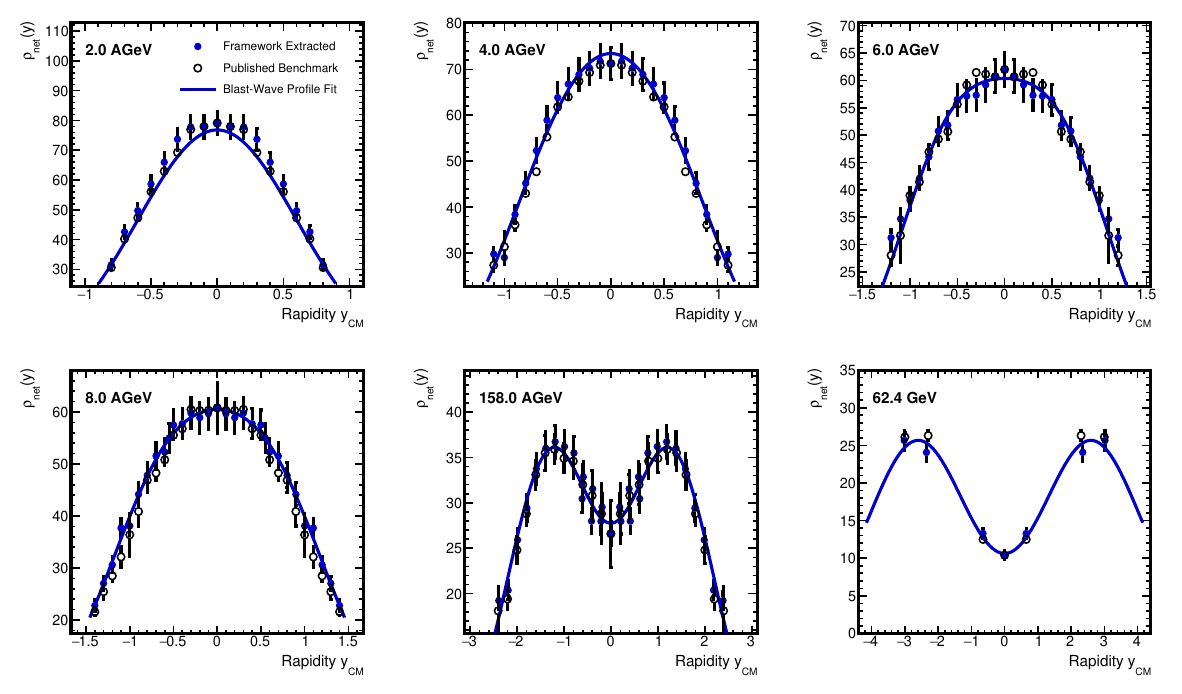}
\caption{\textbf{Extracted net-proton rapidity density profiles.} Solid circles represent values extracted from blast-wave integration; solid curves represent symmetric double-Gaussian fits (Eq.~\eqref{eq:rho_profile}). Open circles show published experimental data~\cite{E895:2001zms,NA49:1998gaz,BRAHMS:2009wlg}.}
\label{fig:rapidity_profiles}
\end{figure*}

We calculate the net-proton rapidity density, $\rho_{\mathrm{net}}(y)$,  by integrating the fitted blast-wave distributions over $p_T$ (see Eq.~\eqref{eq:rho}). The results are plotted in Fig.~\ref{fig:rapidity_profiles}.
The distributions exhibit a transition from a central single-peak structure at AGS energies ($2.0$--$8.0\,A\text{ GeV}$), indicative of strong nuclear stopping, to a double-humped distribution at SPS and RHIC energies ($158\,A\text{ GeV}$ and $\sqrt{s_{\mathrm{NN}}} = 62.4\text{ GeV}$), characterizing the onset of transparency. Our extracted spectra show excellent agreement with published experimental data~\cite{E895:2001zms,NA49:1998gaz,BRAHMS:2009wlg}. This further confirms the reliability of our fitting procedure for the \(p_T\) spectra and validity of the model.

\subsection{Transverse Momentum Broadening}

\begin{figure}[t]
\centering
\includegraphics[width=\columnwidth]{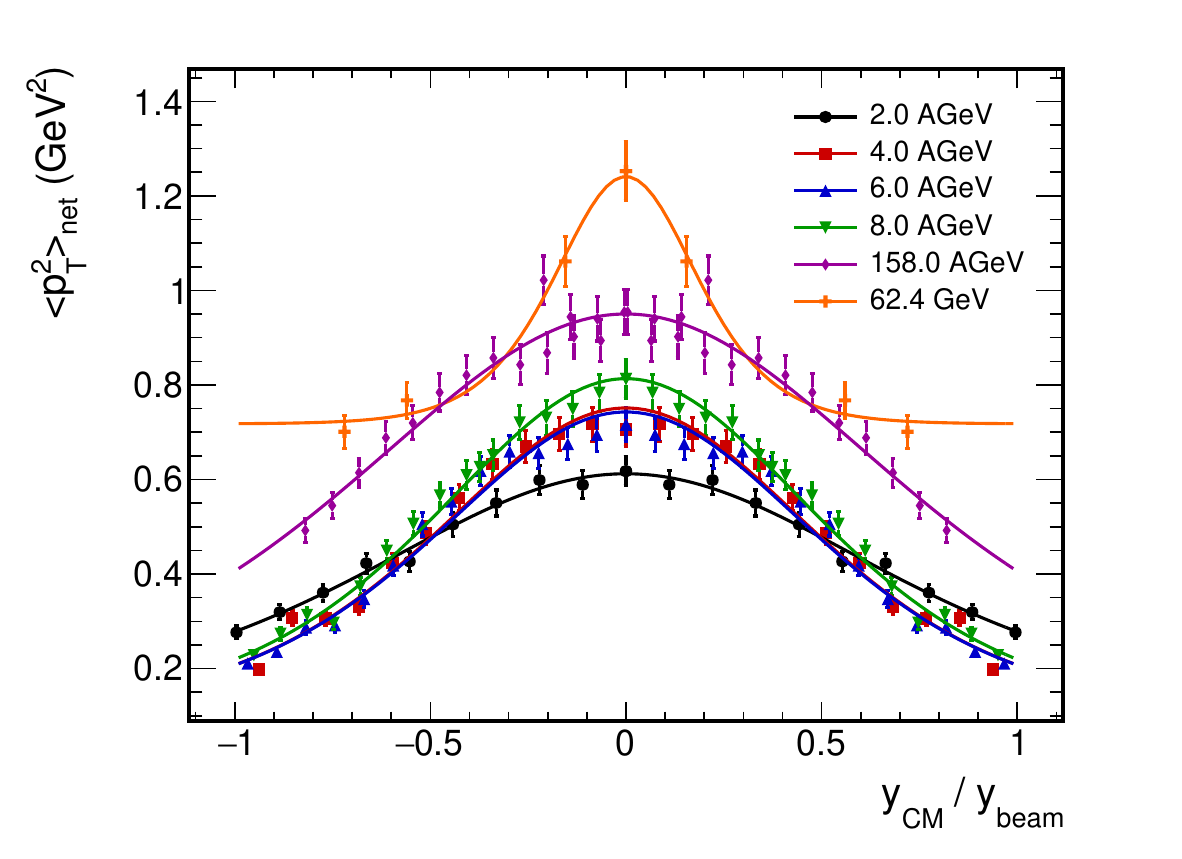}
\caption{\textbf{Normalized transverse momentum broadening.} $\langle p_T^2\rangle_{\mathrm{net}}(y)$ as a function of scaled center-of-mass rapidity $y_{\mathrm{CM}}/y_{\mathrm{beam}}$ across all six beam energies. Solid curves denote parameterizations using the $\operatorname{sech}^2$ functional form (Eq.~\eqref{eq:pt2_profile}).}
\label{fig:pt2_broadening}
\end{figure}

The second moment $\langle p_T^2\rangle_{\mathrm{net}}(y)$ extracted via Eq.~\eqref{eq:pt2} is displayed in Fig.~\ref{fig:pt2_broadening}. The broadening profiles are fitted using Eq.~\eqref{eq:pt2_profile}.

The transition from broad variance profiles at low energies to sharp profiles at high energies physically signals a decoupling of transverse and longitudinal degrees of freedom. This sharpening of the $\langle p_T^2\rangle_{\mathrm{net}}(y)$ profile directly drives the reduction in $\alpha_{\mathrm{bw}}$ via Eq.~\eqref{eq:alpha}, as the gradient term becomes increasingly localized.

\subsection{Excitation Functions of Transport Coefficients}

\begin{figure*}[t]
\centering
\includegraphics[width=\textwidth]{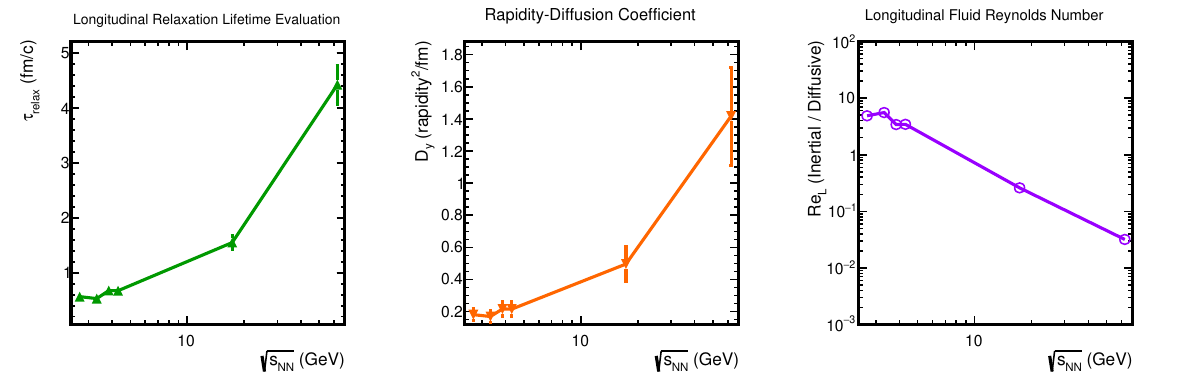}
\caption{\textbf{Extracted transport coefficients as functions of center-of-mass energy $\sqrt{s_{\mathrm{NN}}}$.} (a) Longitudinal relaxation time $\tau_{\mathrm{relax}}$ (in $\text{fm}/c$). (b) Rapidity diffusion coefficient $D_y$ (in $\text{rapidity}^2/\text{fm}$). (c) Dimensionless longitudinal Reynolds number $\mathrm{Re}_L$ on a logarithmic scale, displaying the crossover from advection-dominated ($\mathrm{Re}_L > 1$) to diffusion-dominated ($\mathrm{Re}_L < 1$) dynamics.}
\label{fig:transport_coeffs}
\end{figure*}

Figure~\ref{fig:transport_coeffs} presents the extracted transport coefficients $\tau_{\mathrm{relax}}$, $D_y$, and $\mathrm{Re}_L$ across center-of-mass collision energies. Numerical values are presented in Table~\ref{tab:summary}. It has been observed that the longitudinal relaxation time $\tau_{\mathrm{relax}}$ increases monotonically from $0.566 \pm 0.045\text{ fm}/c$ at $2.0\,A\text{ GeV}$ to $4.429 \pm 0.354\text{ fm}/c$ at $\sqrt{s_{\mathrm{NN}}} = 62.4\text{ GeV}$. This rise signifies a transition from rapid, strong nuclear stopping at low energies to a highly transparent regime at high energies, where the relaxation time approaches the typical fireball lifetime~\cite{Moore:2004tg}. Concurrently, the rapidity diffusion coefficient $D_y$ increases from $0.181 \pm 0.039\text{ rapidity}^2/\text{fm}$ to $1.417 \pm 0.305\text{ rapidity}^2/\text{fm}$. This growth indicates that at higher energies, random momentum kicks from the medium increasingly dominate over directed frictional deceleration.

The longitudinal Reynolds number $\mathrm{Re}_L$ drops by a factor of $\sim 163$ from $\mathrm{Re}_L = 4.88 \pm 0.44$ at $\sqrt{s_{\mathrm{NN}}} = 2.70\text{ GeV}$ to $\mathrm{Re}_L = 0.03 \pm 0.01$ at $62.4\text{ GeV}$. The reduction of $\mathrm{Re}_L$ to $0.03$ at $62.4\text{ GeV}$ provides a macroscopic signature consistent with the decoupling of the baryon junction from valence quarks~\cite{star2026,kharzeev1996,vance1998}, where diffusion overtakes advection as the junction is stranded at mid-rapidity.

\begin{table*}[t]
\caption{Summary of extracted transport parameters, relaxation lifetimes, Reynolds numbers, and stopping fractions across collision energies compared against published values~\cite{thakur2016}.}
\label{tab:summary}
\begin{ruledtabular}
\begin{tabular}{lcd{1.7}d{1.7}d{1.7}d{1.7}d{2.5}d{2.6}}
\multicolumn{1}{c}{System} &
\multicolumn{1}{c}{$\sqrt{s_{\mathrm{NN}}}$ (GeV)} &
\multicolumn{1}{c}{$D_y$ ($\text{rapidity}^2/\text{fm}$)} &
\multicolumn{1}{c}{$\alpha_{\mathrm{bw}}$ ($\text{fm}^{-1}$)} &
\multicolumn{1}{c}{$\tau_{\mathrm{relax}}$ ($\text{fm}/c$)} &
\multicolumn{1}{c}{$\mathrm{Re}_L$} &
\multicolumn{1}{c}{$f_{\mathrm{stop}}$ (\%)} &
\multicolumn{1}{c}{Published $f_{\mathrm{stop}}$ (\%)~\cite{thakur2016}} \\
\hline
$2.0\,A\text{ GeV}$   & 2.70  & 0.181 \pm 0.039 & 1.767 \pm 0.141 & 0.566 \pm 0.045 & 4.88 \pm 0.44 & 59.5 \pm 3.0 & 64.22 \pm 0.10 \\
$4.0\,A\text{ GeV}$   & 3.32  & 0.169 \pm 0.037 & 1.888 \pm 0.151 & 0.530 \pm 0.042 & 5.57 \pm 0.50 & 48.7 \pm 2.4 & 52.16 \pm 0.15 \\
$6.0\,A\text{ GeV}$   & 3.84  & 0.216 \pm 0.047 & 1.479 \pm 0.118 & 0.676 \pm 0.054 & 3.42 \pm 0.31 & 42.3 \pm 2.1 & 48.10 \pm 0.20 \\
$8.0\,A\text{ GeV}$   & 4.31  & 0.215 \pm 0.046 & 1.486 \pm 0.119 & 0.673 \pm 0.054 & 3.45 \pm 0.31 & 39.0 \pm 1.9 & 44.61 \pm 0.12 \\
$158.0\,A\text{ GeV}$ & 17.32 & 0.496 \pm 0.107 & 0.645 \pm 0.052 & 1.551 \pm 0.124 & 0.26 \pm 0.02 & 17.5 \pm 0.9 & 15.25 \pm 2.32 \\
$62.4\text{ GeV}$     & 62.40 & 1.417 \pm 0.305 & 0.226 \pm 0.018 & 4.429 \pm 0.354 & 0.03 \pm 0.00 & 5.9  \pm 0.3 &  4.57 \pm 0.10 \\
\end{tabular}
\end{ruledtabular}
\end{table*}

\subsection{Cross-Validation, Stopping Fractions, and Entropy Production}

\begin{figure*}[t]
\centering
\includegraphics[width=\textwidth]{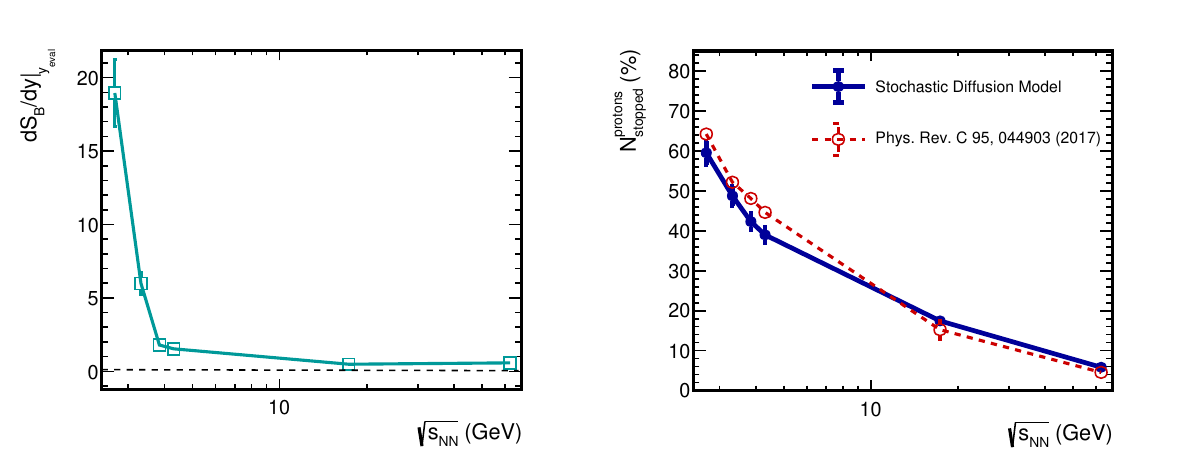}
\caption{\textbf{Stopping fractions and entropy production.}  (a) Baryon stopping entropy production rate $\mathrm{d}S_B/(\mathrm{d}y\,\mathrm{d}\tau)|_{y_{\mathrm{eval}}}$ as a function of center-of-mass energy $\sqrt{s_{\mathrm{NN}}}$. (b) Central stopping fraction $f_{\mathrm{stop}}$ compared against benchmark calculations from Ref.~\cite{thakur2016}. The deviations at low energies arise from the functional forms (two-source vs. double-Gaussian) and the inclusion of radial flow in the blast-wave framework.}
\label{fig:cross_validation}
\end{figure*}

The central stopping fraction $f_{\mathrm{stop}}$, defined as the ratio of the net-baryon yield within $|y| \le 0.5$ to total yield:
\begin{equation}
f_{\mathrm{stop}} = \frac{\int_{-0.5}^{0.5} \rho_{\mathrm{net}}(y)\,\mathrm{d}y}{\int_{-\infty}^{\infty} \rho_{\mathrm{net}}(y)\,\mathrm{d}y} \times 100\%,
\label{eq:fstop_def}
\end{equation}
is compared with calculations from Ref.~\cite{thakur2016} in Fig.~\ref{fig:cross_validation}(b). 

The benchmark values in Table~\ref{tab:summary} are taken from Ref.~\cite{thakur2016}, where the stopping fractions were extracted using a two-source thermal function fitted directly to projected net-proton rapidity distributions. In the present study, stopping fractions are derived from the blast-wave-integrated rapidity density profiles obtained via $p_T$ spectral fitting. 

Examining the excitation function reveals an energy-dependent convergence behavior between both methodologies. In the low-energy regime ($\sqrt{s_{\mathrm{NN}}} \le 5.0\text{ GeV}$), our model predicts a slightly lower stopping fraction than Ref.~\cite{thakur2016} (e.g., $59.5 \pm 3.0\%$ vs. $64.22 \pm 0.10\%$ at $2.0\,A\text{ GeV}$). This difference arises from two physical and methodological factors. First, the two-source function in Ref.~\cite{thakur2016} assumes localized thermal fireballs shifted by $\pm y_s$ scaled by hyperbolic cosines ($\cosh(y/y_s)$), which concentrates baryon density tightly around mid-rapidity. In contrast, our symmetric double-Gaussian parameterization incorporates continuous stochastic diffusion broadening into the tails, slightly reducing the fraction contained strictly within $|y| \le 0.5$. Second, the blast-wave model explicitly accounts for collective radial flow ($\beta \approx 0.32$), which modifies the $p_T$ spectrum. The conversion of compressive energy into transverse collective expansion broadens the effective phase space, redistributing a small fraction of net-protons beyond $|y| = 0.5$.

In the intermediate to high-energy regime ($\sqrt{s_{\mathrm{NN}}} \ge 17.3\text{ GeV}$), agreement between both methods is reasonable ($17.5 \pm 0.9\%$ vs. $15.25 \pm 2.32\%$ at $158\,A\text{ GeV}$, and $5.9 \pm 0.3\%$ vs. $4.57 \pm 0.10\%$ at $62.4\text{ GeV}$). In this transparent regime, the fragmentation peaks separate widely from mid-rapidity ($y_{\mathrm{beam}} \gg 1$), and central baryon density is governed entirely by stochastic diffusion and topological junction transport rather than the geometry of projectile remnants. 

Figure~\ref{fig:cross_validation}(a) depicts the stopping entropy production rate $\mathrm{d}S_B/(\mathrm{d}y\,\mathrm{d}\tau)|_{y_{\mathrm{eval}}}$, which exhibits a sharp decline from low to high collision energies, reflecting the high dissipation of early-stage compression. This behavior is consistent with the observation that baryon stopping dynamics are directly relevant to total entropy production in heavy-ion collisions ~\cite{Steinberg:2006bhw}. Furthermore, the rate perfectly mirrors the kinematic crossover observed in the Reynolds number. When the system is advection-dominated at low energies, the intense frictional drag of the incoming baryons violently transfers ordered longitudinal kinetic energy into random thermal motion, generating massive entropy. As the collision energy increases and the system transitions into the diffusion-dominated regime ( $\mathrm{Re}_L < 1$), the medium becomes highly transparent; the transport is then governed by weak stochastic scattering~\cite{Denicol:2018wdp}, resulting in minimal thermodynamic dissipation. Therefore, the simultaneous collapse of both $\mathrm{Re}_L$ and the entropy production rate establishes a unified, consistent thermodynamic picture: the baryon current transitions from a highly dissipative, frictionally stopped state at low energies to a nearly dissipationless, transparent state at high energies.

\subsection{Implications for Baryon Junctions and Future Facilities}
\label{sec:implications}

The extracted transport coefficients provide crucial macroscopic constraints on the microscopic mechanisms of baryon transport. Standard valence-quark string-deceleration models predict a gradual decrease in stopping efficiency \cite{Mehtar-Tani:2008hpn}. In stark contrast, the topological baryon junction model \cite{kharzeev1996, vance1998, kopeliovich1999} suggests that baryon number is carried by a Y-shaped gluon configuration, predicting non-zero transfer over large rapidity intervals. Our observation of a sharp crossover from advection-dominated (\(\mathrm{Re}_L > 1\)) to diffusion-dominated (\(\mathrm{Re}_L < 1\)) dynamics, alongside the steep rise in \(\tau_{\mathrm{relax}}\), quantitatively supports this junction mechanism. As the medium becomes more transparent at higher energies, the steep rise in \(\tau_{\mathrm{relax}}\) indicates that the junction decouples from the valence quarks, suppressing multi-scattering stopping in favor of stochastic diffusion.

This macroscopic picture is strongly supported by recent STAR measurements in isobar (\(\mathrm{Ru}+\mathrm{Ru}\) vs. \(\mathrm{Zr}+\mathrm{Zr}\)) and photonuclear (\(\gamma+\mathrm{Au}\)) collisions \cite{star2026}. The observed \(B/\Delta Q\) ratio exceeding valence-quark expectations strongly supports the picture that baryon number is transported by the Y-shaped gluon configuration rather than by valence quarks, which is highly consistent with the transport dynamics extracted in this work.

This cross-validated framework offers a clear, falsifiable baseline for upcoming high-density experiments, including CBM at FAIR~\cite{Heuser:2009gg,ablyazimov2017}, MPD at NICA ~\cite{Kekelidze:2017ual}, and the CEE facility~ \cite{MA:2020oqt}. If the baryon junction mechanism dominates at intermediate energies, we predict a significant suppression of the diffusion coefficient \(D_y\) and an extension of \(\tau_{\mathrm{relax}}\) compared to a purely valence-quark scenario. Conversely, if valence-quark scattering dominates, the extracted \(\mathrm{Re}_L\) should continue a smooth monotonic decline without the sharp advective-to-diffusive crossover observed here \cite{wolschin2007}. Therefore, future high-precision measurements from these facilities will provide a decisive experimental test of baryon transport in the high-\(\mu_B\) QCD phase diagram \cite{stephanov1998, ratti2021}.

\section{SUMMARY}
\label{sec:summary}

We have applied an integrated analysis pipeline to extract non-equilibrium transport properties of baryon stopping from net-proton $p_T$ spectra across beam energies from $E_{\mathrm{lab}} = 2.0\,A\text{ GeV}$ to $\sqrt{s_{\mathrm{NN}}} = 62.4\text{ GeV}$.

By simultaneously analyzing the net-proton rapidity density $\rho_{\mathrm{net}}(y)$ and transverse broadening $\langle p_T^2\rangle_{\mathrm{net}}(y)$ within a stochastic Fokker-Planck framework, we extracted the longitudinal friction coefficient $\alpha_{\mathrm{bw}}$, rapidity diffusion coefficient $D_y$, relaxation time $\tau_{\mathrm{relax}}$, and longitudinal Reynolds number $\mathrm{Re}_L$. The longitudinal Reynolds number demonstrates a transition from $\mathrm{Re}_L \sim \mathcal{O}(10)$ at low energies to $\mathrm{Re}_L < 1.0$ at top energies, quantitatively mapping the boundary between inertial baryon stopping and diffusive transparency. This collapse of $\mathrm{Re}_L$ quantitatively characterizes the transport transition expected from the decoupling of the baryon junction from valence quarks, strongly supporting the topological transport mechanism. Extracted stopping fractions agree with established theoretical benchmarks. This unified dataset provides calibrated baseline inputs for forthcoming experiments mapping high-density matter at FAIR, NICA and CEE.

Finally, we note that our analysis is deliberately bounded at \(\sqrt{s_{\mathrm{NN}}} = 62.4\text{ GeV}\). Extending this framework to top RHIC energies (\(\sqrt{s_{\mathrm{NN}}} = 200\text{ GeV}\)) is mathematically ill-conditioned rather than physically necessary. At \(200\text{ GeV}\), the net-proton yield at mid-rapidity approaches zero due to the near-exact cancellation of protons and anti-protons. Because our transport formalism relies on the net-baryon density \(\rho_{\mathrm{net}}(y)\) as the fundamental variable, the logarithmic gradient \(\partial \ln \rho_{\mathrm{net}}/\partial y\) diverges as \(\rho_{\mathrm{net}} \to 0\), making the extraction of \(\alpha_{\mathrm{bw}}\) and entropy production highly unstable. Since the transition to the fully transparent, diffusion-dominated regime (\(\mathrm{Re}_L \ll 1\)) is already completely reached at \(62.4\text{ GeV}\), including higher-energy data would not alter the physical conclusions. Instead, it would merely inflate the systematic uncertainties. Therefore, the \(62.4\text{ GeV}\) bound ensures the most robust and mathematically sound baseline for mapping the stopping-to-transparency crossover.

\end{document}